\documentclass[epj]{webofc}
\usepackage[utf8]{inputenc}
\usepackage[varg]{txfonts}   
\usepackage{booktabs}
\usepackage{amssymb}
\usepackage{xcolor}
\definecolor{darkred}{rgb}{0.4,0.0,0.0}
\definecolor{darkgreen}{rgb}{0.0,0.4,0.0}
\definecolor{darkblue}{rgb}{0.0,0.0,0.4}
\usepackage[bookmarks,linktocpage,colorlinks,
    linkcolor = darkred,
    urlcolor  = darkblue,
    citecolor = darkgreen]{hyperref}

\usepackage{subfigure}
\wocname{EPJ Web of Conferences}
\woctitle{Lattice2017}
\begin{document}
%
\selectlanguage{english}
\reversemarginpar{DESY 17-165}
\title{%
Simulation of an ensemble of $N_f=2+1+1$ twisted mass clover-improved fermions at physical quark masses
}
\author{%
\firstname{Jacob} \lastname{Finkenrath}\inst{1}\fnsep\thanks{Speaker, \email{j.finkenrath@cyi.ac.cy}} \and
\firstname{Constantia} \lastname{Alexandrou}\inst{1,2} \and
\firstname{Simone}  \lastname{Bacchio}\inst{2,3}\and
\firstname{Panagiotis} \lastname{Charalambous}\inst{1,2}\and \\
\firstname{Petros} \lastname{Dimopoulos}\inst{4,5}\and 
\firstname{Roberto} \lastname{Frezzotti}\inst{4}\and
\firstname{Karl} \lastname{Jansen}\inst{5}\and  
\firstname{Bartosz} \lastname{Kostrzewa}\inst{6}\and \\
\firstname{Giancarlo} \lastname{Rossi}\inst{4,5}\and 
\firstname{Carsten} \lastname{Urbach}\inst{6}
}
\institute{%
Computation-based Science and Technology Research Center, The Cyprus Institute, Nicosia,  20, Constantinou Kavafi Str., Nicosia 2121, Cyprus
\and
Department of Physics, University of Cyprus, P.O. Box 20537,  Nicosia 1678, Cyprus
\and
Bergische Universit{\"a}t Wuppertal, Wuppertal, Germany
\and 
Dipartimento di Fisica, Università di Roma ``Tor Vergata" and INFN,
Sezione di Roma 2, Via della Ricerca Scientifica - 00133 Rome, Italy
\and
Centro Fermi - Museo Storico della Fisica e Centro Studi e Ricerche
Enrico Fermi, Compendio del Viminale, Piazza del Viminiale 1, I-00184, Rome, Italy
\and
NIC, DESY, Zeuthen, Germany
\and 
HISKP (Theory), Rheinische Friedrich-Wilhelms-Universit{\"a}t Bonn, Bonn, Germany
}

\abstract{%
We present a general strategy aimed at generating $N_f=2+1+1$ configurations with quarks at their physical mass using maximally 
twisted mass fermions to ensure automatic $O(a)$ improvement, in the presence of a clover term tuned to
reduce the charged to neutral pion mass difference.
The target system, for the moment, is a lattice of size
$64^3 \times 128$ with a lattice spacing $a\sim 0.08$~fm.
We show preliminary results on the pion and kaon mass and decay constants. 
}
\maketitle

\section{Introduction}\label{intro}

During the last decade, lattice QCD simulations with light quarks having physical masses on large enough volumes have 
become feasible owing to improvements in lattice actions and numerical algorithms 
as well as due to steadily increasing computational power.
In this work, we illustrate a strategy aimed at generating $N_f=2+1+1$ configurations
with quarks at the physical mass using maximally twisted fermions to ensure automatic $O(a)$ improvement.
This bonus does not come without a price, as the twisted regularization breaks isospin leading to an
undesired (negative) neutral pion to charged mass splitting. It has been shown in~\cite{Abdel-Rehim:2015pwa} 
that a suitably tuned clover term can ease the problem and allow for stable simulations at 
affordable lattice spacings and volumes (sec.~\ref{sec:1}). 
The tuning procedure is quite 
complicated as a number of parameters (bare quark masses, critical mass, and the clover term coefficient,
$c_{SW}$) have to be simultaneously set at appropriate values.
In sec.~\ref{sec:2} we discuss a viable strategy that allows with relative small effort and good signal-to-noise ratio to desired tuning 
in a way that weakly depends on the light quark sector.
Going to large lattice volumes, as needed to simulate
pions with close-to-physical mass, leads to an increased
density of very small Dirac operator eigenvalues, which is expected
to stabilise the MD behaviour of HMC-like algorithms but also requires 
the use of very well preconditioned and efficient linear solvers.
We will discuss the practical aspects of our Monte Carlo 
simulation strategy and related issues in sec.~\ref{sec:3}. Finally in sec.~\ref{sec:4}, we will present results 
on the pion and kaon masses and decay constants using the $N_f =2+1+1$ ensemble.

\section{Setup}
\label{sec:1}

The action used in our simulation is given by
\begin{equation}
 S = \beta \cdot S_g \;-\; \chi_\ell^\dagger \{ D^\dagger(\mu_\ell)D(\mu_\ell)\}^{-1} \chi_\ell \;-\; \chi_h^\dagger \{D_{1+1}^\dagger(\mu_\sigma,\mu_\delta) D_{1+1}(\mu_\sigma,\mu_\delta) \}^{-1/2} \chi_h
\end{equation}
with $S_g$ the Iwasaki improved gauge action \cite{Iwasaki:1985we}, $\beta$ the bare coupling constant
and $\chi_\ell$ and $\chi_h$ pseudofermion fields.
The light quark-doublet is made of the up- and down-quarks,  described in the isospin symmetric limit by a degenerated twisted mass
operator with twisted mass parameter $\mu_\ell$. The heavy quark doublet is made by the strange and charm quarks
and are described by the mass non-degenerated twisted mass operator given by
  \begin{equation}
   D_{1+1}(\mu_\sigma,\mu_\delta) =  D_{W}(\kappa,c_{sw}) \otimes {1} + i \mu_\sigma (\gamma_5 \otimes \tau_3) - \mu_\delta (1 \otimes \tau_1) = \begin{bmatrix}
                                                                            D_W + i \gamma_5 \mu_\sigma & -\mu_\delta \\
                                                                            -\mu_\delta  & D_W - i \gamma_5 \mu_\sigma
                                                                           \end{bmatrix} 
\end{equation}  
with bare mass parameter $\mu_\sigma$ and $\mu_\delta$. The Wilson Dirac operator $D_{W}(\kappa,c_{sw})$
depends on the bare mass parameter $\kappa = \frac{1}{2 (4+m) }$. A Sheikholeslami--Wohlert-term \cite{Sheikholeslami:1985ij}
is added. For degenerated masses one gets that
$ \det\{ D^\dagger(\mu_\ell)D(\mu_\ell)\} = \det\{ D_{1+1} (\mu_\ell, 0)\}$. 

In the twisted mass discretization of lattice QCD all the correlators that are
expectation values of parity even multilocal operator products are by
symmetry free of linear lattice artefacts if the Partially Conserved Axial
Current (PCAC) mass  vanishes, which is equivalent to tuning the bare mass parameter $\kappa(m)$
to its critical value \cite{Frezzotti:2003ni,Frezzotti:2003xj}. In this case, the renormalized light quark mass is at maximal
twist and related to $m_\ell = \mu_\ell/Z_P$ with $Z_P$ the pseudoscalar renormalization constant.
This   fine tuning of $\kappa(m) \rightarrow \kappa_{crit}(m)$
 will be discussed in the next section. 
 
The twisted mass Wilson-Dirac operator corresponding to the light quark sector
is protected from zero eigenmodes for a finite twisted mass term $\mu_\ell >0$.
In fact $D^\dagger(\mu_\ell) D(\mu_\ell) = D_{W}^\dagger D_{W} + \mu_\ell^2$ 
with $\min\{ \lambda \} \geq \mu_\ell$.
From a numerical point of view this holds the promise to be able to reach small quark masses
with $\mu_\ell>0$ including the physical value.

However, the lattice regularization breaks some continuum symmetries
that are recovered after extrapolating to zero lattice spacing.
Besides the breaking of the chiral symmetry due to the Wilson term,
a finite twisted mass term breaks isospin symmetry by introducing a mass-splitting in the pion triplet. 

The neutral to charge pion mass splitting can be estimated in chiral perturbation theory.
To first order one finds~\cite{Sharpe:2004ny}
\begin{equation}
  (m_{\pi^0}^2 - m_{\pi^\pm}^2) \approx  4 \cdot c_2 \cdot a^2 
\end{equation}
where $c_2$ depends on low energy constants and it is found to
be negative for twisted mass fermions~\cite{Herdoiza:2013sla}. This means
that the neutral pion mass may vanish for finite light quark masses
triggering a phase transition into  a non-physical phase
induces by lattice artefacts. 

\begin{figure}[thb] 
\centering
\includegraphics[width=6.5cm,clip]{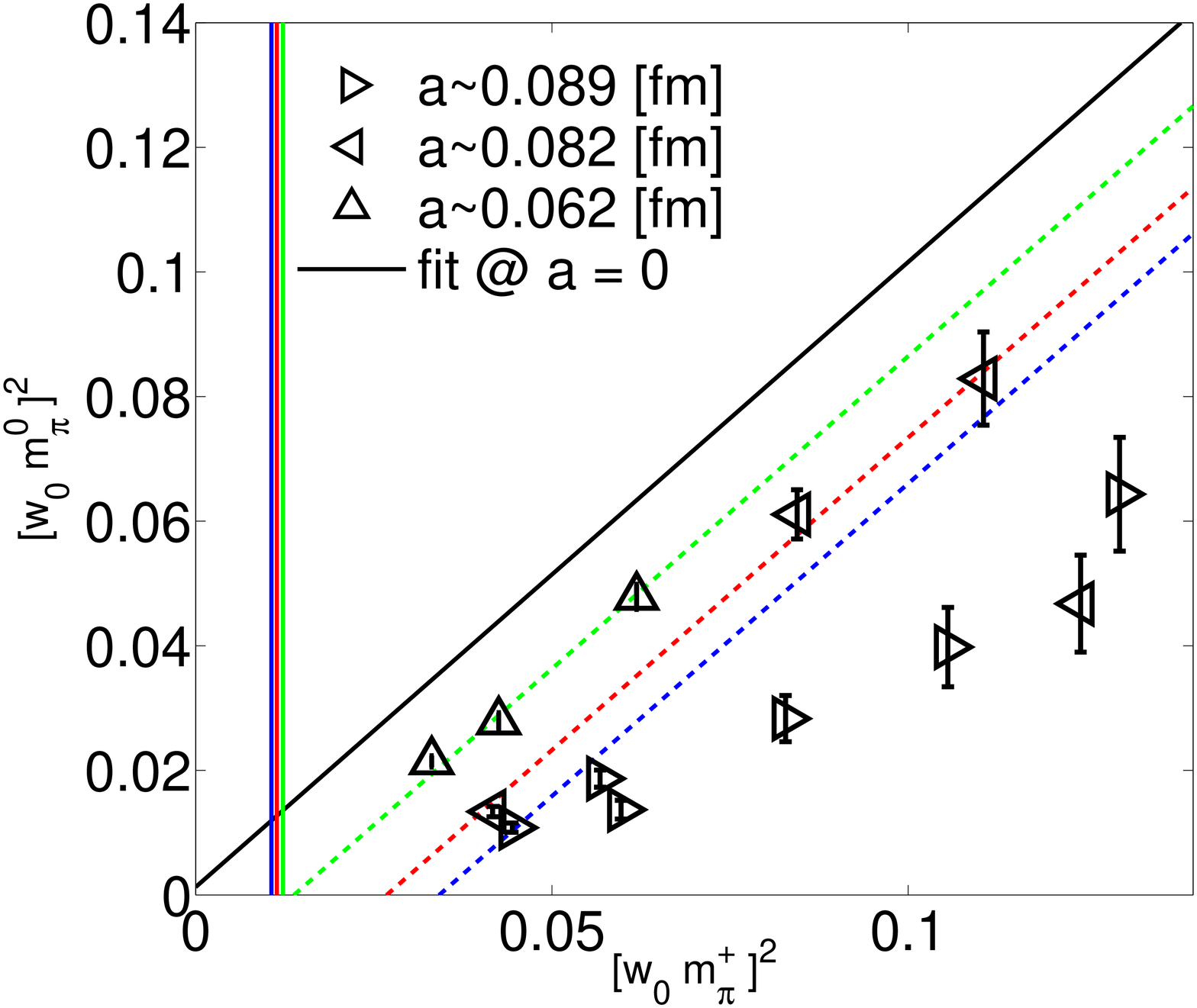}
\includegraphics[width=6.5cm,clip]{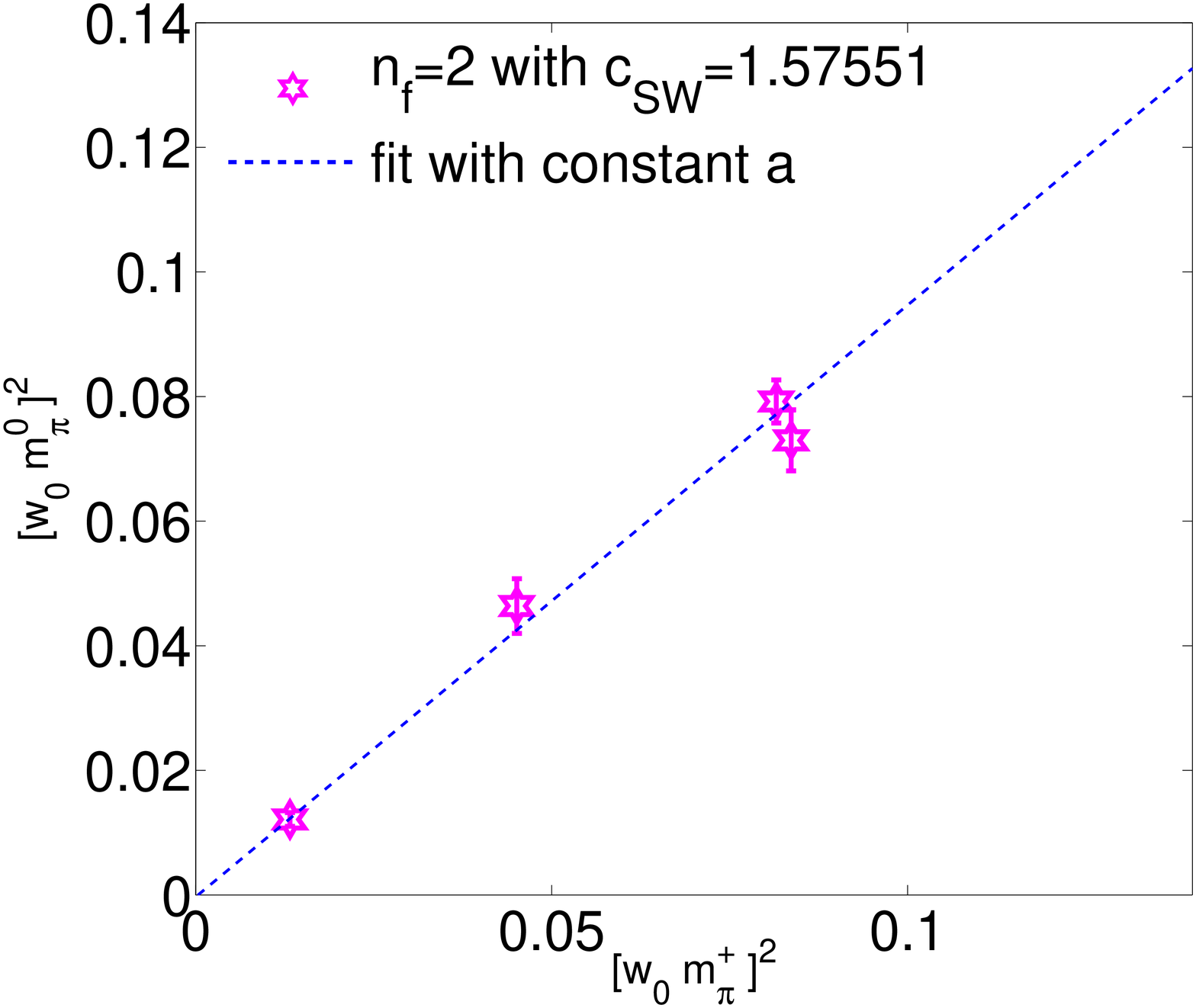}
\vspace{-0.3cm}
\caption{ { Left:} The squared neutral pion mass is plotted against the squared charged pion mass in units of $w_0$ 
for $N_f=2+1+1$ ETMC ensembles with no clover term i.e.~taking $c_{SW}=0$. We use a generalized fit excluding ensembles with $m_\pi > 400 \; \textrm{MeV}$ and show results for  three lattice spacings and extrapolating to  $a=0$. The line
corresponds to the physical value of the charged pion mass. 
We used the definition of \cite{Borsanyi:2012zs} for $w_0$.
{ Right:}  The neutral pion mass  for the $N_f=2$ ensemble with a clover term.
}
\label{fig:isospin}
\end{figure}

In Fig.~\ref{fig:isospin} we show the isospin mass splitting between the charged
and neutral pion. For $N_f = 2+1+1$ twisted mass fermion ensembles with $c_{SW} = 0$ the mass splitting between the neutral and charged pions
is sizable
for the three different lattice spacings examined.  It follows that for lattice spacing $a\gtrsim 0.05 \; \textrm{fm}$ 
the neutral pion mass would vanish for physical light quark masses making simulation
at the physical point impossible. However, it was shown that by adding a clover term the isospin breaking effects can be suppressed~\cite{Abdel-Rehim:2015pwa}.
This was demonstrated in the case of  the $N_f = 2$ twisted mass fermion  ensembles with $c_{SW} = 1.57551$ and a lattice spacing
of $a = 0.0938 \; \textrm{fm}$.
In particular, for a charged pion mass of $m_\pi = 130 \; \textrm{MeV}$ the pion mass difference between the neutral and charge pions 
is smaller than $ \sqrt{| m_{\pi^0}^2 - m_{\pi^\pm}^2 |}  \lesssim 6 \; \textrm{MeV}$
and simulation of twisted mass fermions at physical light quarks are 
possible for $a<0.1 \; \textrm{fm}$.
   
\vspace{-0.4cm}  
   
\vspace{-0.3cm}
\section{Tuning towards physical quark masses}
\label{sec:2}

We start by discussing the tuning procedure towards physical quark mass, 
using the $N_f=2+1+1$ twisted mass fermion discretization at a target lattice spacing of $a\sim 0.8 \; \textrm{fm}$.
The inclusion in the action of a clover term suppresses isospin violation effects.
Using an estimate from 1--loop tadpole boosted perturbation theory given by
$c_{SW} \cong 1+0.113 g^2_{plaq}$ \cite{Aoki:1998qd} with 
$g^2_{plaq} = g_0^2/P$ and $P$ the plaquette expectation value.
This yields $c_{SW} = 1.69$.
The tuning procedure consists of two major parts: i) tuning towards the critical mass and ii)
 tuning  the heavy quarks towards their physical values. The two steps are strongly correlated.
To tune to critical mass we use  the vanishing of the  PCAC mass, which depends on the light valence quarks
and to  tune the strange and charm quark masses we use dimensionless rations 
depending only on the strange and charm valence quarks.

\begin{table}[thb]
  \centering
  \begin{tabular}{ccccccc }
    \hline \hline \\[-2.0ex]
  V               & $\beta$        &    $\mu_{\ell}$   & $\mu_\sigma$  & $\mu_\delta$              &  $\kappa$     &   $c_{SW}$             \\[0.8ex]
    \hline \\[-1.8ex]
 $128\times64^3$ &   1.778         &  0.00072     &   0.1246864      & 0.1315052                           &  $0.1394265$   &   $1.69$      \\[0.8ex]
    \hline \hline 
  \end{tabular}
  \caption{The table shows the parameter used in the simulations of the target ensemble with pion masses close 
  to the physical point.\label{tab:parametersL64}}
\end{table}

\vspace{-0.3cm}
For the tuning towards the critical mass we impose \cite{Boucaud:2008xu}
\begin{equation}
 \frac{Z_A m_{PCAC}(\kappa)}{\mu_\ell} \lesssim 0.1
 \label{eq:PCACcriter}
\end{equation}
with $Z_A$ the axial renormalization constant roughly given by $\sim 0.8$. For the tuning
of the heavy quark masses, we introduce an additional step
by setting the mass parameters $\mu_s$ and $\mu_c$ of the Osterwalder Seiler fermions such that
\begin{equation}
 A_1 = \frac{\mu_c}{\mu_s} \equiv 11.8 \qquad \textrm{and} \qquad B_2 =  \frac{m_{D_s}}{f_{D_s}} \equiv 7.9 ~.
\end{equation}
The final step is to match the masses of the Kaons involving an Osterwalder Seiler
and a non-degenerated twisted mass valence strange quark, respectively.
In this way the matching of valence and sea quark mass parameters is 
independent of finite size effects and with a negligibly small dependence
on the light quark mass entering only through sea quark effects.

\begin{figure}[thb] 
\centering
\includegraphics[width=5.6cm,clip]{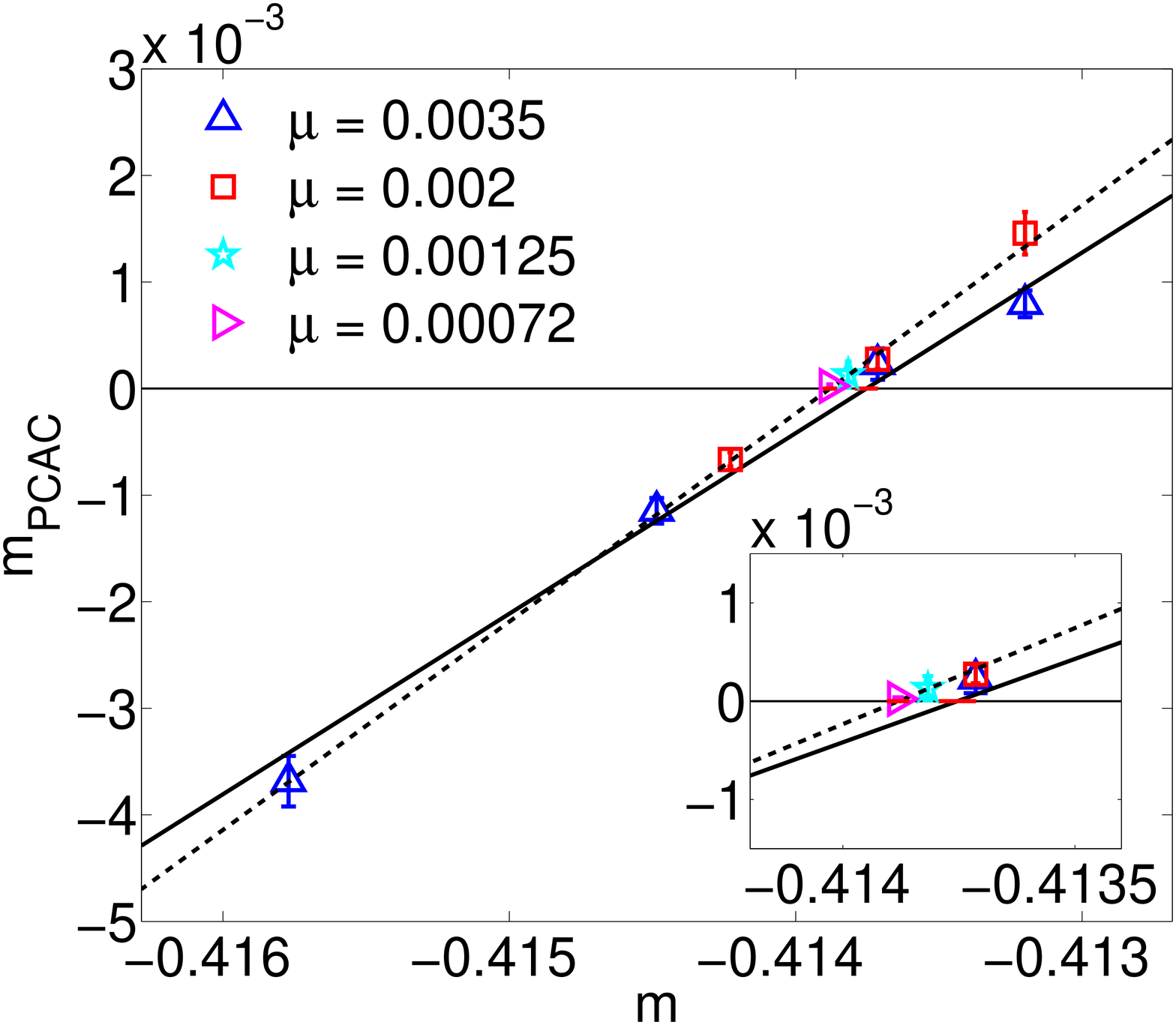}
\includegraphics[width=6.4cm,clip]{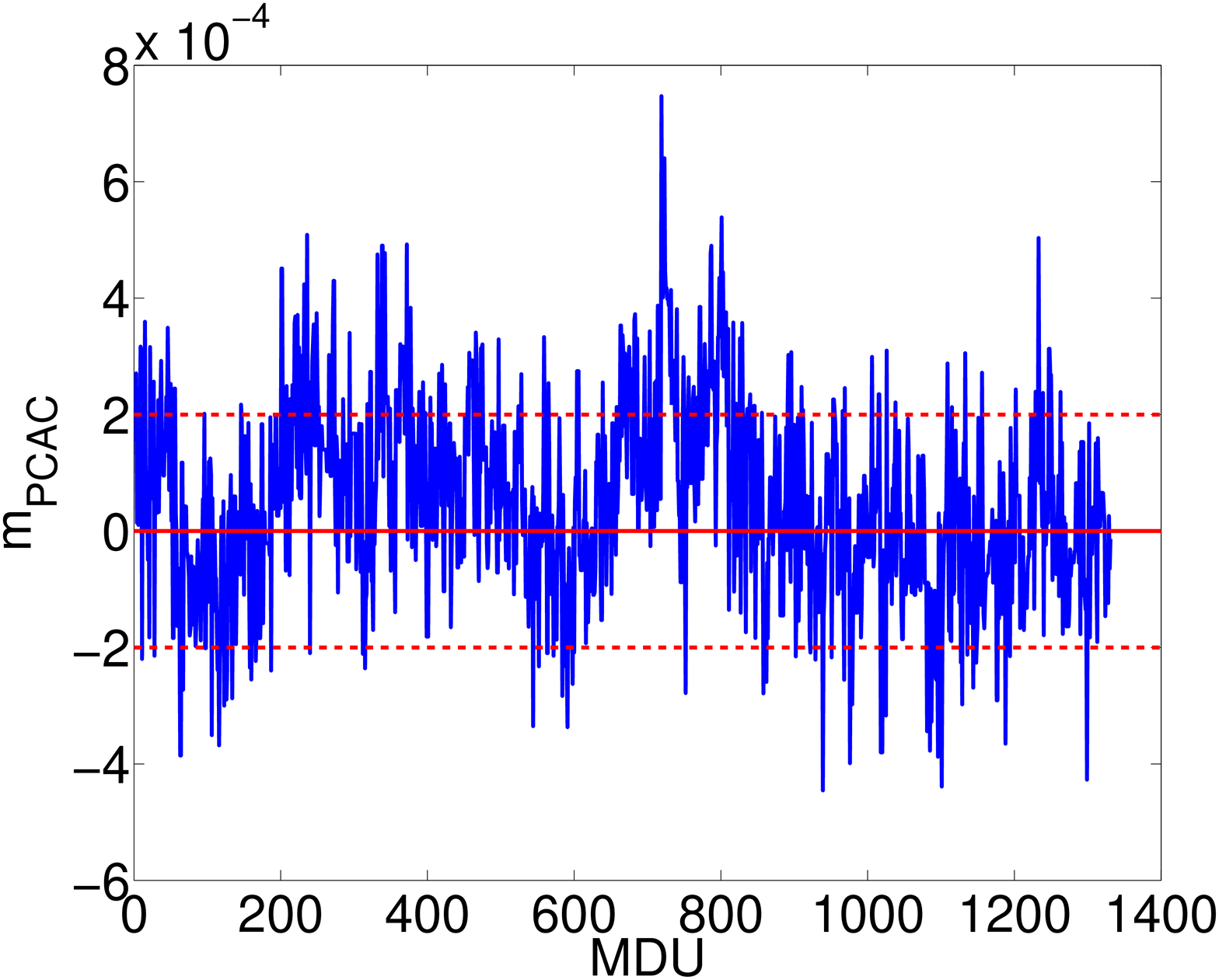}
\vspace{-0.3cm}
\caption{Left: PCAC mass versus $\kappa$ in the tuning.
Blue triangles are for the lattices $N_s=24$, the red  symbols for the lattices $N_s=32$, the cyan star for a lattice of $N_s=32$ with
a pion mass of $m_\pi \sim 180 \; \textrm{MeV}$ and the magenta point corresponds to the target ensemble with $N_s=64$. 
Right: The history of the PCAC mass of our target lattice  in units of molecular dynamics (MDU).}
\label{fig:tunPCAC}
\end{figure}

The approach was as follows
\begin{itemize}
 \item  We start with simulation using  lattices of spatial extent $N_s=24$ at four different mass values $\kappa$ and
	 $m_\pi \sim 310 \; \textrm{MeV}$. The parameters $\beta$ and $\mu_\sigma,\, \mu_\delta$
	 are initially estimated by a set of preliminary studies aimed at investigating their interdependencies~\cite{Kostrzewa2017Maximally}. 
	 Using a linear fit for the PCAC mass as a function  of the bare light quark mass parameter 
	 we estimate $\kappa_{crit}$ as shown in Fig.~\ref{fig:tunPCAC}) and recheck the value of $c_{sw}$ and the lattice spacing $a$ by using Wilson flow observables.
  \item With the estimated critical mass value we simulate a larger lattice size with spatial extent $N_s=32$ and we check and re-tune the heavy
	quark masses by calculating the properties of the kaon and D-mesons.
  \item Based on the re-tuned heavy quark parameters $\mu_\sigma$ and $\mu_\delta$
	we run three simulations again with spatial lattice size $N_s=32$ at different $\kappa$ values and 
	a pion mass of $m_\pi \sim 230 \; \textrm{MeV}$.
        Again we estimate the critical mass by using a linear fit to the PCAC mass for the generated ensembles as shown in Fig.~\ref{fig:tunPCAC}.
        We then cross-check the estimated value by simulating an ensemble with a pion mass $m_\pi \sim 180 \textrm{MeV}$.
  \item Finally we simulate  our target lattice size with $N_s=64^3$ and temporal extent $N_T=128$ using pion masses close to the physical value. 
\end{itemize}

\vspace{-0.3cm}
\vspace{-0.3cm}
The PCAC masses of the target ensemble are shown Fig.~\ref{fig:tunPCAC}
and final tuned parameters are found in Table~\ref{tab:parametersL64}.
On the first 1200 MDUs we find that the average
PCAC mass satisfies $|m_{PCAC}| <  0.00005$, which 
fulfills the bound given in Eq.~(\ref{eq:PCACcriter}) that ensures sufficient suppression of the linear lattice artefacts.

\section{Stable algorithms}
\label{sec:3}
  
\begin{figure}[thb] 
\centering
\includegraphics[width=5cm,clip]{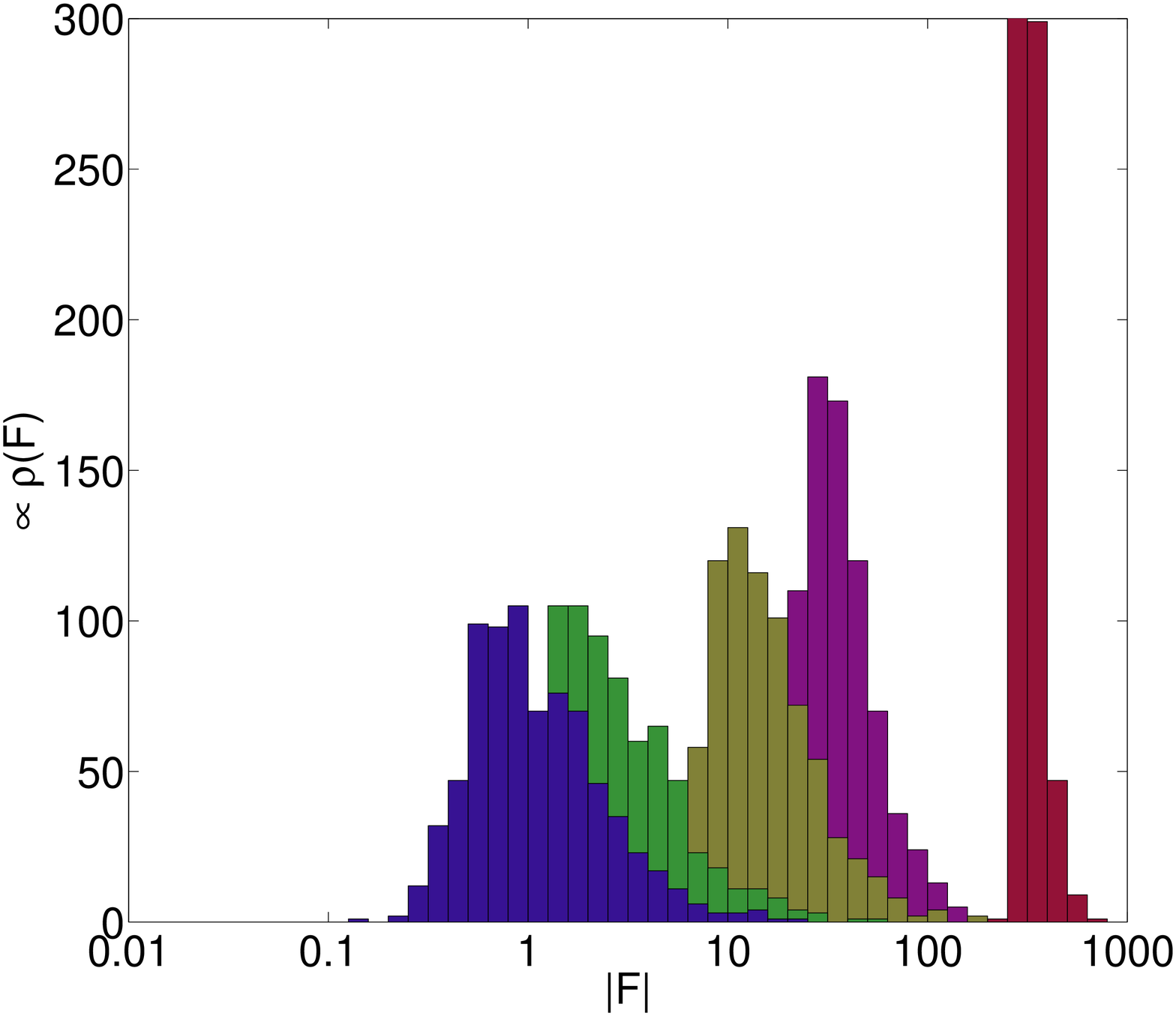}
\includegraphics[width=5cm,clip]{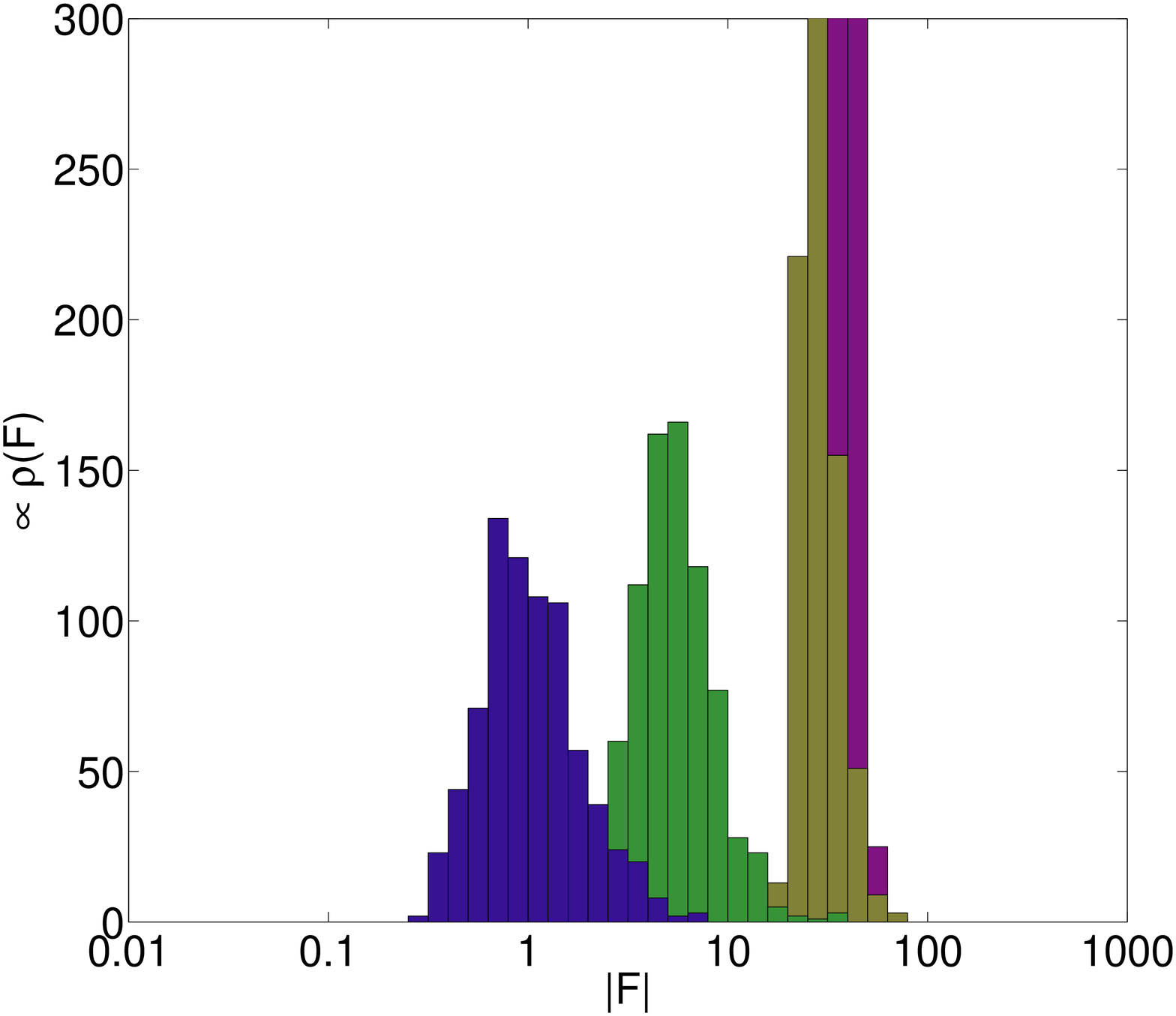}
\caption{Left: Distribution of the maximal norm in the force terms  of the light quark sector, where the force terms depending on the largest Hasenbusch mass-shift
are shown in blue for $\rho=0.003$, in green for $\rho=0.0012$, in yellow for $\rho=0.01$ and in purple
for $\rho=0.1$. The force term for the full operator is shown in red.
Right: The Distribution of the maximal norm
of the force terms, which are involved in the
rational approximation. The distribution
in blue corresponds to the terms $i=9, 10$, in green to the
terms $i=7, 8$, in yellow to the terms $i=6, 5, 4$ and in purple to the terms $i=3, 2, 1$ 
in the rational approximation. The smallest shift is given by $i=10$.}
\label{fig:forceterms}
\end{figure}

We use the Hybrid Monte Carlo (HMC) algorithms \cite{Duane:1987de}
with Hasenbusch mass-preconditioning \cite{Hasenbusch:2002ai} for the light quark sector 
and the rational HMC \cite{Clark:2006fx} for the heavy quark sector. 
The software package \textit{tmLQCD} \cite{Jansen:2009xp} with the multi-grid solver DDalphaAMG \cite{Bacchio:2016bwn} is employed.
For the inversions of the light and heavy quark sector 
a three-level DD-$\alpha$AMG solver \cite{Frommer:2013fsa,Alexandrou:2016izb}, and even-odd preconditioned conjugate 
gradient (CG) solver and a multi-mass shifted CG solver are used (see Ref.~\cite{Bacchio:2017} for more details). The gauge fields are integrated by
using a 6-level nested minimal norm scheme of second order.
We show the distribution of the maximal force terms
for the light and heavy quark sectors in Fig.~\ref{fig:forceterms}.
The rational approximation is done for the interval
$[0.000014 \;\; 1]$ by involving 10 terms, while a correction
term is included in the acceptance step.
Using 12 integration steps on the outer level and 384 on the inner level,
the integration with length $\tau=1$ yields a stable simulation,
where the maximal energy violation is given by $\textrm{max}(\delta H) = 5.67$
with an acceptance ratio of $77 \%$.
\begin{figure}[thb] 
\centering
\includegraphics[width=6cm,clip]{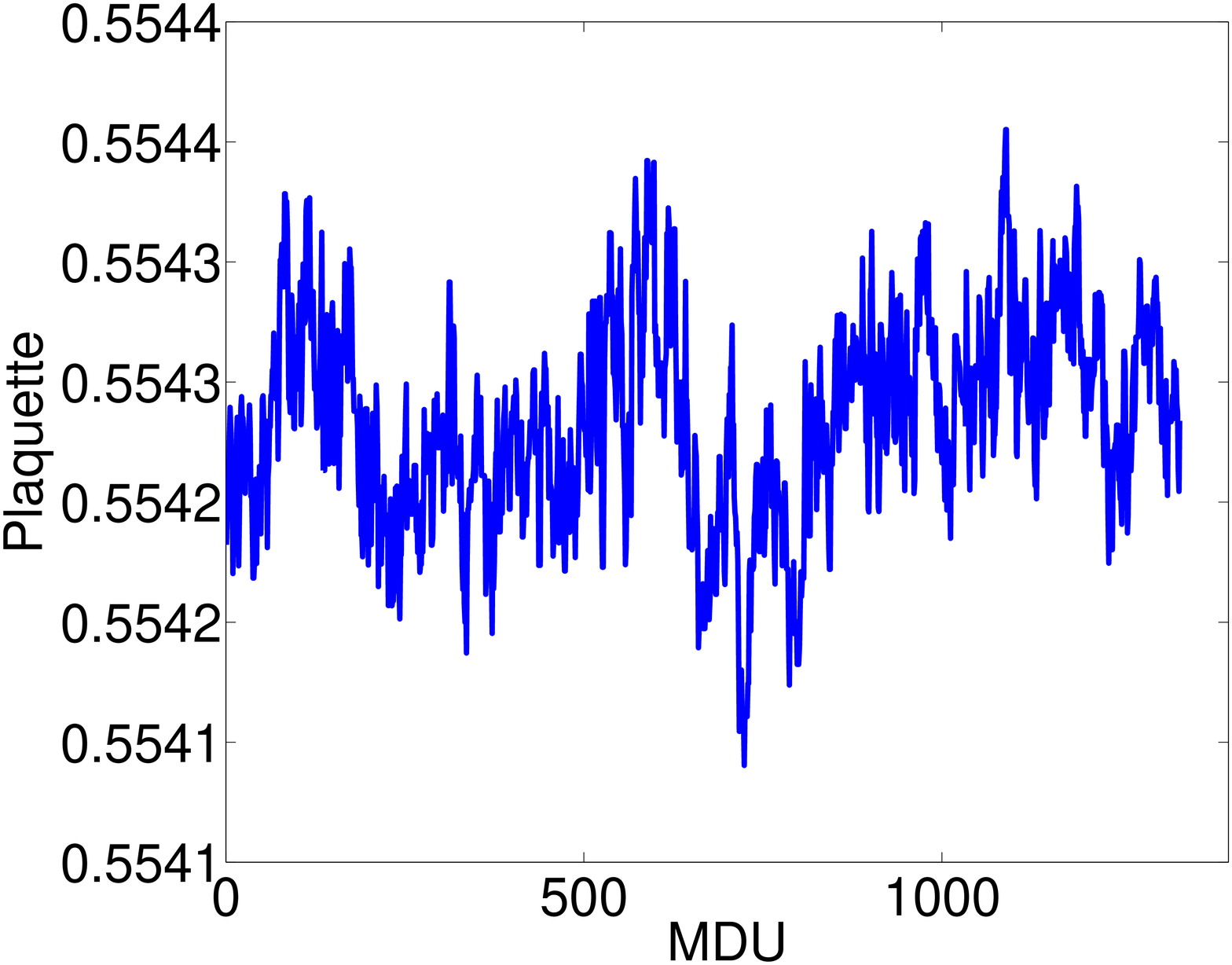}
\includegraphics[width=6cm,clip]{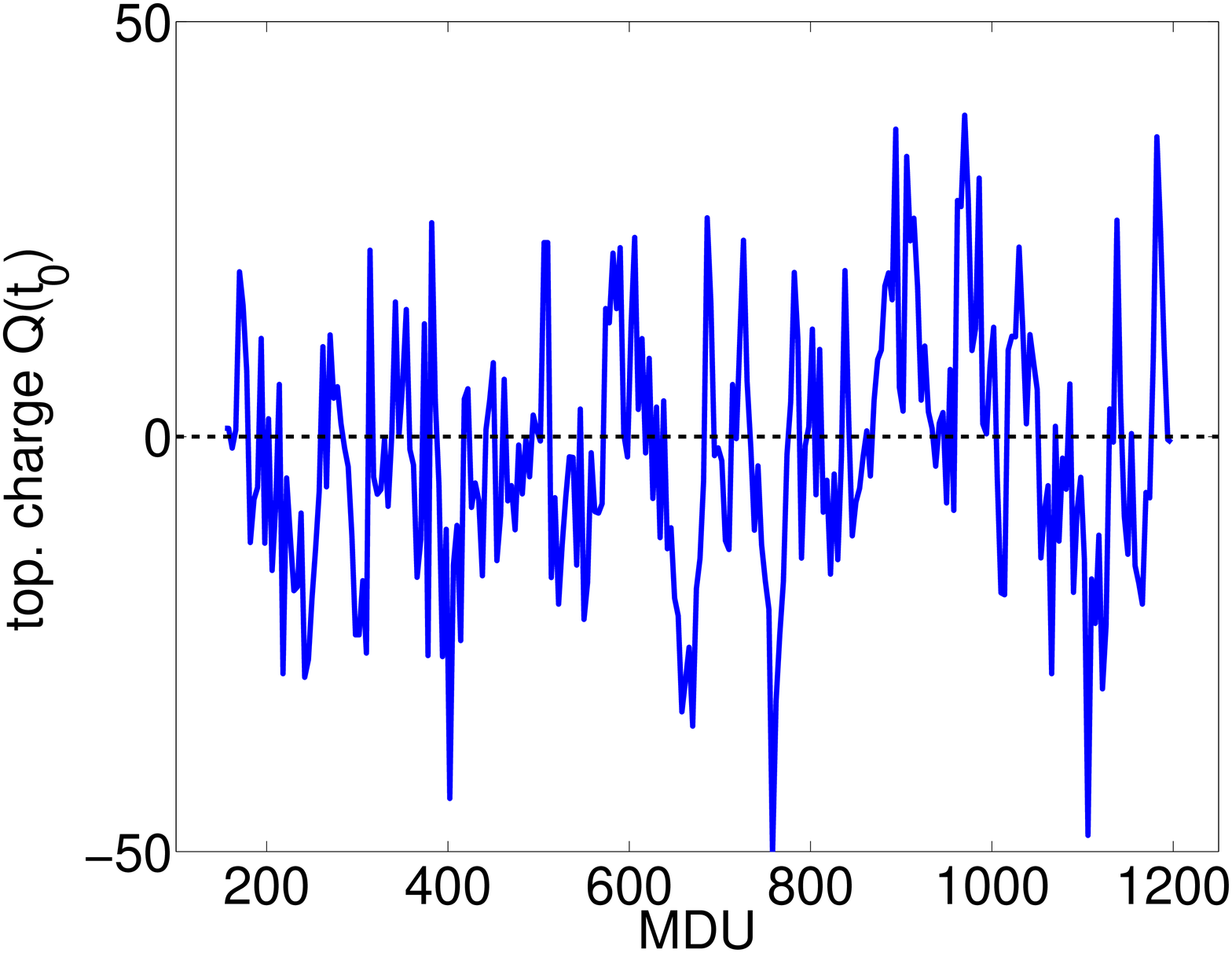}
\caption{{Left:} The history of the plaquette.
{Right:} The history of the topological charge defined using the gauge expression
at gradient flow time $t=t_0$, is shown.}
\label{fig:platopo}
\end{figure}

In the  case of the plaquette it can be  seen that the use of the Hasenbusch mass-preconditioning 
yields a quite stable run. As shown in Fig.~\ref{fig:platopo} the
history of the plaquette shows some larger oscillations for MDU$<800$.
This provides an estimate for the integrated autocorrelation time 
$\tau_{int}$. We find $\tau_{int}= 31(13)$. Similar oscillations are found in the
history of the PCAC mass, also shown in Fig.~\ref{fig:tunPCAC}. They are seen to be 
anticorrelated with the plaquette oscillations. Nevertheless, we find that other observables,
such as the topological charge,
are not affected, and shows frequent oszillations. 
As shown in Fig.~\ref{fig:platopo} the algorithm 
can efficiently sample the topological charge with an autocorrelation
time of $\tau_{int} = 8(3) $.

\section{Observables}
\label{sec:4}

\begin{figure}[thb] 
\centering
\includegraphics[width=7cm,clip]{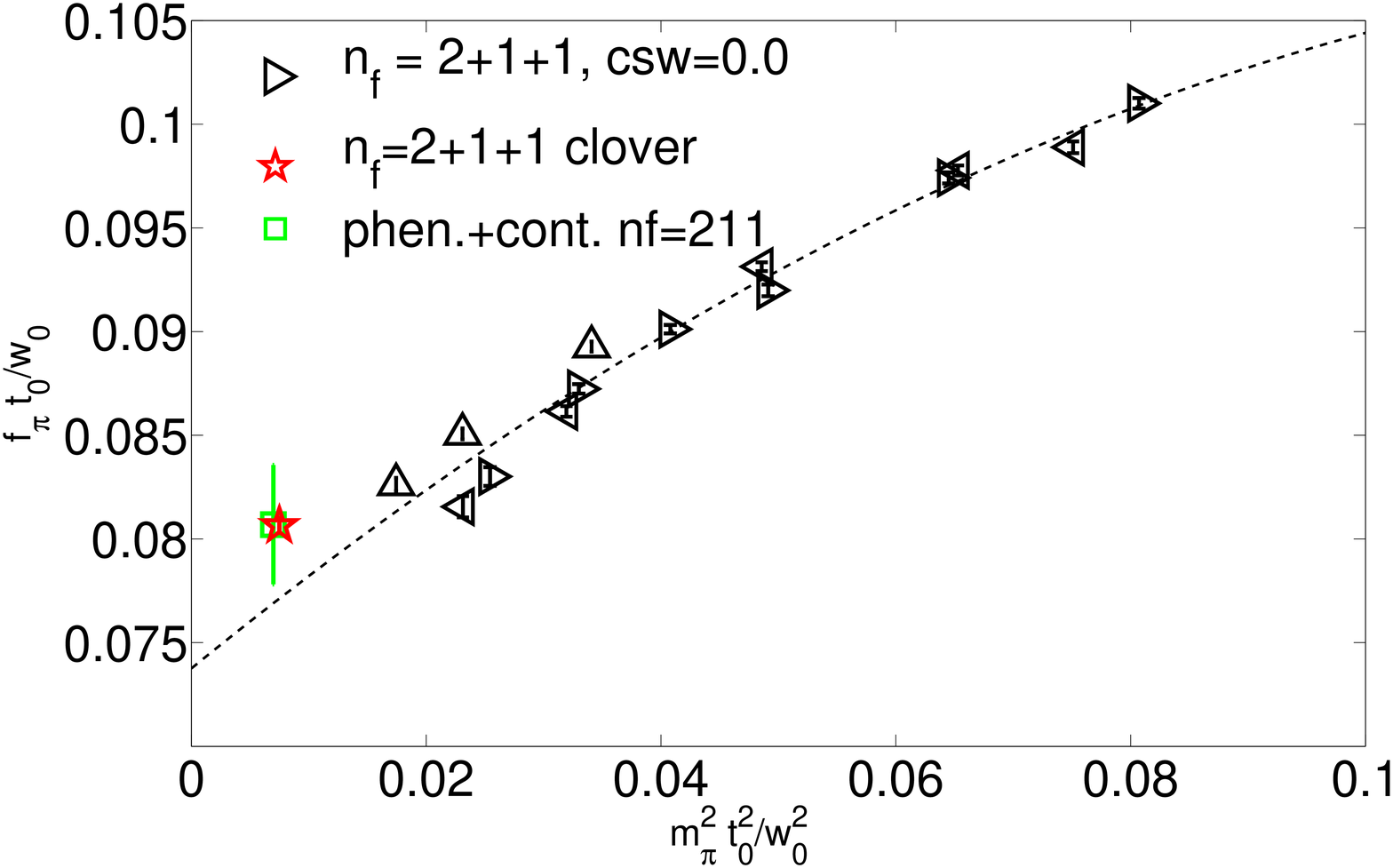}
\includegraphics[width=7cm,clip]{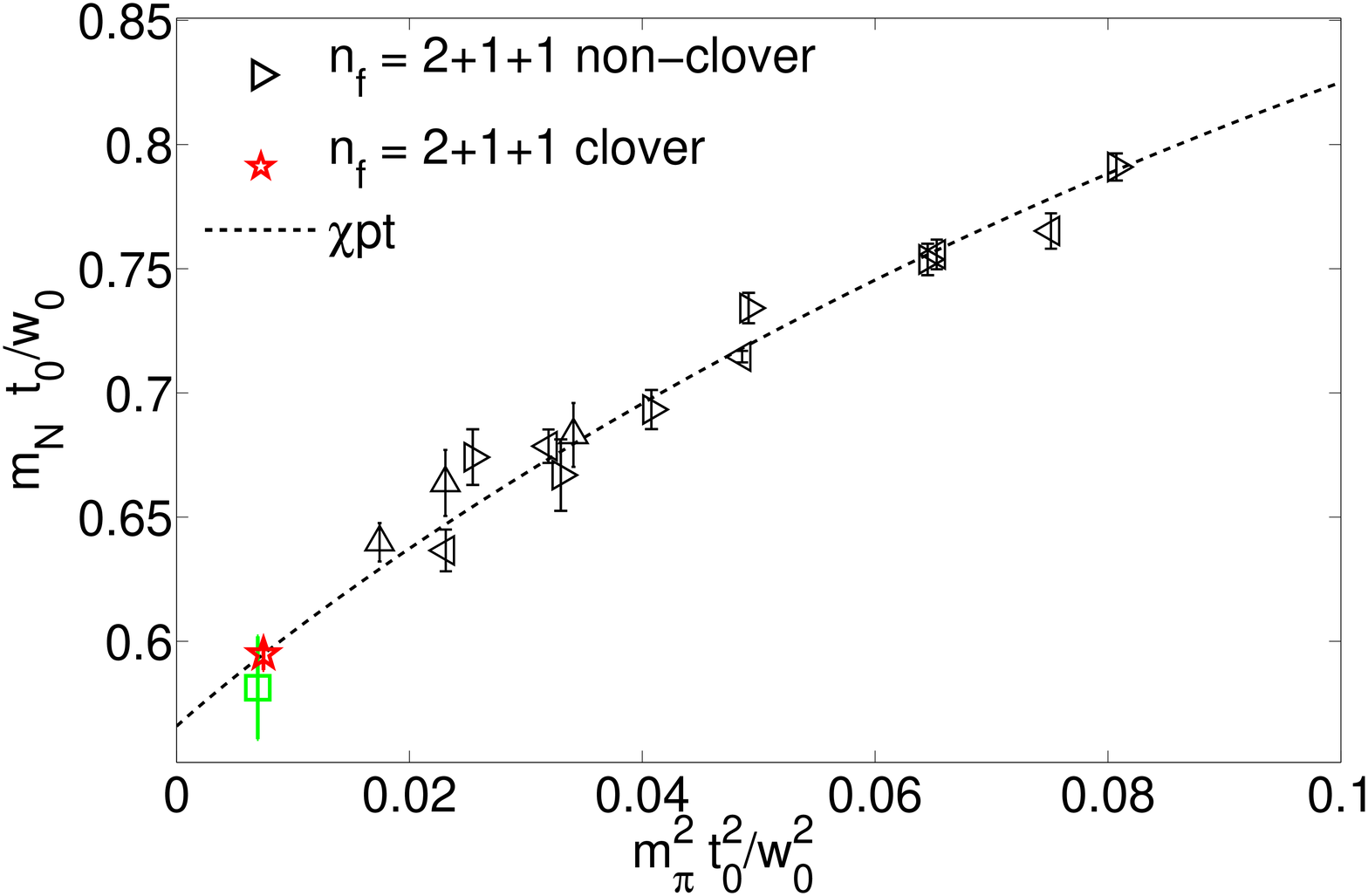}
\caption{ Left: The pion decay constant $f_\pi$ is plotted against the pion mass in units of $t_0/w_0 a$.
We show results for  $N_f=2+1+1$ ensembles without a clover term for three different values of $\beta$ (triangles, black), The red star shows 
the pion decay constant of  our target ensemble with $c_{SW}=1.69$. 
The continuum value is shown by the green square. Right: We give results on the nucleon mass $m_N$
from our target ensemble (red  star) compared with results obtained using $N_f=2+1+1$ twisted mass ensembles 
without a clover term (open triangles).
DIfferent triangle orientations correspond to different lattice spacings
with $a\sim 0.89 \,\textrm{fm}$ for the orientation ``to the right'', $a\sim 0.82 \,\textrm{fm}$ for ``to the left'' and
$a\sim 0.67 \,\textrm{fm}$ for ``up'' .}
\label{fig:fpi}
\end{figure}

In this section, we show preliminary results
extracted from two-point functions using  our
target ensemble of tab.~\ref{tab:parametersL64}. See Ref.\cite{Alexandrou:2017} for a detailed discussion on how these measurements are performed.
In the left panel of Fig.~\ref{fig:fpi} we show the pion mass dependence of the pion decay constant $f_\pi$
in comparison with results obtained using  $N_f=2+1+1$ twisted mass fermions
with $c_{SW}=0$ and smallest pion mass of around $m_\pi\sim 210 \; \textrm{MeV}$ in  units of  the gradient flow observables $t^2_0/w^2_0 a^2$.
The results are fitted to a quadratic function in $m^2_\pi$ to guide
the eye. Note that finite size effects were not corrected. The measured pion decay constant of our target ensemble  agrees with the continuum value, which was determined using
continuum extrapolations of the gradient flow observables determined in\cite{Borsanyi:2012zs} and a
pion decay constant of $130.2 \; \textrm{MeV}$.

In the right panel of Fig.~\ref{fig:fpi} we show the nucleon mass $m_N$ in units of $t^2_0/w^2_0 a^2$
as a function of the  charged pion mass squared $m_\pi^2$. We fit the results
using  $N_f=2+1+1$ twisted mass fermion ensembles  without 
a clover term with the help of leading order chiral perturbation theory~\cite{Gasser:1987rb,Tiburzi:2008bk}. 
The agreement between the chirally extrapolated value with the value obtained using our current physical $N_f=2+1+1$ ensemble indicates small cut-off
effects for the nucleon mass in agreement with previous observations~\cite{Alexandrou:2017xwd}.

\begin{figure}[thb] 
\centering
\includegraphics[width=6cm,clip]{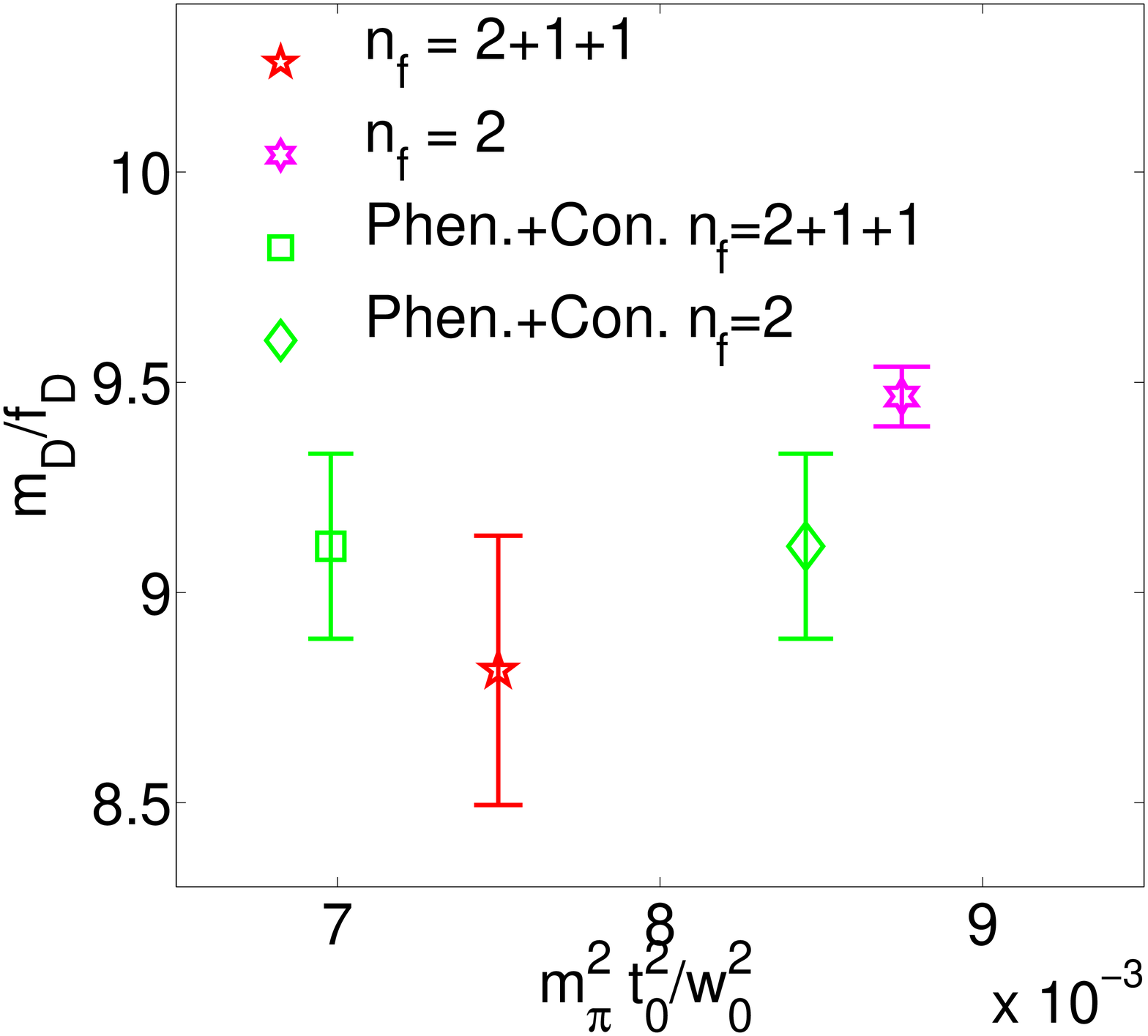}
\includegraphics[width=6cm,clip]{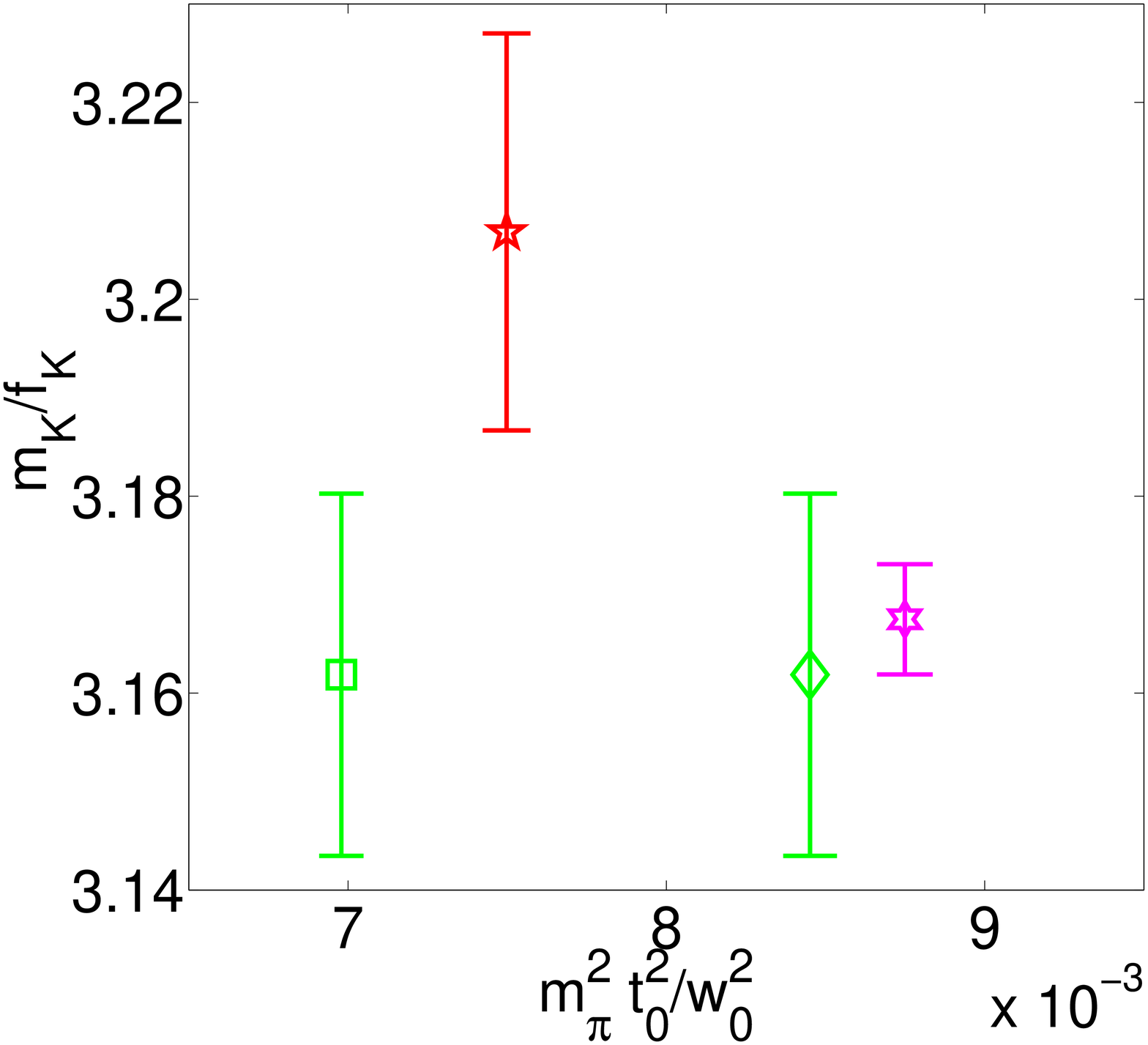}
\caption{{Left:} The ratio of the decay constant of the D-meson over its mass for the $N_f=2+1+1$ ensemble (red star) and for
 the $N_f=2$ twisted mass clover ensemble with a  pion mass of $m_\pi = 130 \; \textrm{MeV}$ (magenta star). The green symbols 
 correspond to the continuum points, using continuum extrapolated gradient flow observables $t_0$ and $w_0$ for
 $n_f=2+1$ in \cite{Borsanyi:2012zs} (square) and for the $N_f=2$ in \cite{Bruno:2013gha} (diamond).
{ Right:} The  ratio of the decay constant of the kaon  over its mass as a function of
the pion mass in units of $t_0^2/w_0^2$ with the same symbols as those used for the D-meson.}
\label{fig:mDmKmfp}
\end{figure}

In the heavy quark sector, we use our tuned values of the strange and charm quarks to evaluate the decay constant of the kaon and D-mesons.
In Fig.~\ref{fig:mDmKmfp}  we show the ratio of the pseudoscalar mass over the 
pseudoscalar decay constant as a function of the pion mass and compare it to
the value obtained using the 
physical $N_f=2$ ensemble with a clover term~\cite{Abdel-Rehim:2015pwa}.
The ratio involving the D-meson agrees with the expectation 
while the ratio by the Kaon is above the physical value. 
A possible reason for this behavior could be the tuning conditions 
which primarily affects the charm quark mass possibly yielding large cut-off effects.

\section{Conclusion}

In this work we illustrate how it is posible to set up stable simulations for  
$N_f=2+1+1$ twisted mass fermions with a clover term to maximal twist and approximately physical values of the light, strange and charm quark masses. 
Introducing a clover
term help to supress isospin breaking effects in the pion triplet and makes the simulation within the twisted mass fermion approach at the physical point feasible.
Our tuning conditions for
fixing the heavy quark masses and  the light bare quark
mass to  maximal twisted are presented. Moreover, we show that, by using 
Hasenbusch--mass preconditioning combined with the RHMC, simulations are stable and that the force terms are under control.
Using the $N_f=2+1+1$ ensemble we show results on the pion decay constant,
the nucleon mass and the ratios of the pseudoscalar mass over the decay constant~\cite{Alexandrou:2017}.
We find a  nucleon to pion mass ratio $m_N/m_\pi= 6.87(7) $  and  kaon and D-meson mass ratio $m_K/m_\pi=3.58(2)$ and  $m_D/m_\pi=13.2(9)$. 
Additional we listed pseudoscalar meson masses and their ratio with their decay constant in tab.~\ref{tab:mesons}.
The lattice spacing as determined from the nucleon mass  turns out to be $a\sim 0.082(2) \; \textrm{fm}$.

\begin{table}[thb]
  \centering
    \caption{In the table are listed the masses and the decay constants of the charged pseudoscalar mesons.}
  \begin{tabular}{|rl|rl|rl|}
     \hline 
     \vspace{-0.2cm} & & & & & \\
   $a m_\pi$       &    0.0567(2)      &  $a m_K$   &    0.2031(9) &  $a m_D$     &   0.748(7)    \\[0.8ex]
   $m_\pi/f_\pi$   &    1.074(5)      &  $m_K/f_K$   &   3.206(20) &  $m_D/f_D$   &   8.81(32)  \\[0.8ex]
    \hline 
  \end{tabular}\label{tab:mesons}
\end{table}

\subsection*{Acknowledgments}
We thank the members of the ETM collaboration for helpful and inspiring discussions. Additional thanks goes
to U.~Wenger for providing the gradient flow data for the $N_f=2+1+1$ ensembles with $c_{SW}=0$.
J. F. thanks  PRACE Fourth Implementation
Phase (PRACE-4IP) program of the European Commission under grant agreement No 653838 for financial support.
This project has received funding from the  Horizon 2020 research and innovation program of the European Commission
under the Marie Sklodowska-Curie grant agreement No 642069.
Part of this work was supported by the Sino-German CRC110.
The authors gratefully acknowledge the Gauss Centre for Supercomputing e.V.~for funding the project \emph{pr74yo} by 
providing computing time on the GCS Supercomputer SuperMUC at Leibniz Supercomputing Centre.

\bibliography{lattice2017}

\begin{thebibliography}{24}

\bibitem{Abdel-Rehim:2015pwa}
A.~Abdel-Rehim et~al. (ETM), Phys. Rev. \textbf{D95}, 094515 (2017),
  \texttt{1507.05068}

\bibitem{Iwasaki:1985we}
Y.~Iwasaki, Nucl. Phys. \textbf{B258}, 141 (1985)

\bibitem{Sheikholeslami:1985ij}
B.~Sheikholeslami, R.~Wohlert, Nucl. Phys. \textbf{B259}, 572 (1985)

\bibitem{Frezzotti:2003ni}
R.~Frezzotti, G.C. Rossi, JHEP \textbf{08}, 007 (2004),
  \texttt{hep-lat/0306014}

\bibitem{Frezzotti:2003xj}
R.~Frezzotti, G.C. Rossi, Nucl. Phys. Proc. Suppl. \textbf{128}, 193 (2004),
  [,193(2003)], \texttt{hep-lat/0311008}

\bibitem{Sharpe:2004ny}
S.R. Sharpe, J.M.S. Wu, Phys. Rev. \textbf{D71}, 074501 (2005),
  \texttt{hep-lat/0411021}

\bibitem{Herdoiza:2013sla}
G.~Herdoiza, K.~Jansen, C.~Michael, K.~Ottnad, C.~Urbach, JHEP \textbf{05}, 038
  (2013), \texttt{1303.3516}

\bibitem{Borsanyi:2012zs}
S.~Borsanyi et~al., JHEP \textbf{09}, 010 (2012), \texttt{1203.4469}

\bibitem{Aoki:1998qd}
S.~Aoki, R.~Frezzotti, P.~Weisz, Nucl. Phys. \textbf{B540}, 501 (1999),
  \texttt{hep-lat/9808007}

\bibitem{Boucaud:2008xu}
P.~Boucaud et~al. (ETM), Comput. Phys. Commun. \textbf{179}, 695 (2008),
  \texttt{0803.0224}

\bibitem{Kostrzewa2017Maximally}
B.~Kostrzewa, Ph.D. thesis, Humboldt-Universität zu Berlin,
  Mathematisch-Naturwissenschaftliche Fakultät (2017)

\bibitem{Duane:1987de}
S.~Duane, A.D. Kennedy, B.J. Pendleton, D.~Roweth, Phys. Lett. \textbf{B195},
  216 (1987)

\bibitem{Hasenbusch:2002ai}
M.~Hasenbusch, K.~Jansen, Nucl. Phys. \textbf{B659}, 299 (2003),
  \texttt{hep-lat/0211042}

\bibitem{Clark:2006fx}
M.A. Clark, A.D. Kennedy, Phys. Rev. Lett. \textbf{98}, 051601 (2007),
  \texttt{hep-lat/0608015}

\bibitem{Jansen:2009xp}
K.~Jansen, C.~Urbach, Comput. Phys. Commun. \textbf{180}, 2717 (2009),
  \texttt{0905.3331}

\bibitem{Bacchio:2016bwn}
S.~Bacchio, C.~Alexandrou, J.~Finkenrath, A.~Frommer, K.~Kahl, M.~Rottmann, PoS
  \textbf{LATTICE2016}, 259 (2016), \texttt{1611.01034}

\bibitem{Frommer:2013fsa}
A.~Frommer, K.~Kahl, S.~Krieg, B.~Leder, M.~Rottmann, SIAM J. Sci. Comput.
  \textbf{36}, A1581 (2014), \texttt{1303.1377}

\bibitem{Alexandrou:2016izb}
C.~Alexandrou, S.~Bacchio, J.~Finkenrath, A.~Frommer, K.~Kahl, M.~Rottmann,
  Phys. Rev. \textbf{D94}, 114509 (2016), \texttt{1610.02370}

\bibitem{Bacchio:2017}
S.~Bacchio, et~al., \textbf{LATTICE2017} (2017)

\bibitem{Alexandrou:2017}
C.~Alexandrou et~al., in preparation  (2017)

\bibitem{Gasser:1987rb}
J.~Gasser, M.E. Sainio, A.~Svarc, Nucl. Phys. \textbf{B307}, 779 (1988)

\bibitem{Tiburzi:2008bk}
B.C. Tiburzi, A.~Walker-Loud, Phys. Lett. \textbf{B669}, 246 (2008),
  \texttt{0808.0482}

\bibitem{Alexandrou:2017xwd}
C.~Alexandrou, C.~Kallidonis, Phys. Rev. \textbf{D96}, 034511 (2017),
  \texttt{1704.02647}

\bibitem{Bruno:2013gha}
M.~Bruno, R.~Sommer (ALPHA), PoS \textbf{LATTICE2013}, 321 (2014),
  \texttt{1311.5585}

\end{thebibliography}

\end{document}